\documentstyle[sprocl]{article}

\def\Journal#1#2#3#4{{#1} {\bf #2}, #3 (#4)}

\def\NPB{{\em Nucl. Phys.} B}
\def\PLB{{\em Phys. Lett.}  B}
\def\PRL{\em Phys. Rev. Lett.}
\def\PRD{{\em Phys. Rev.} D}
\def\ZPC{{\em Z. Phys.} C}
\def\PTP{\em Prog. Theor. Phys.}

\def\bsll{$b \rightarrow s \ell^+ \ell^- $ }
\def\bxsll{$B \rightarrow X_s \ell^+ \ell^- $ }
\def\bxsee{B \rightarrow X_s e^+ e^-  }
\def\bxsmm{B \rightarrow X_s \mu^+ \mu^-  }
\def\bxstt{B \rightarrow X_s \tau^+ \tau^- }

\newcommand{\ra}{\rightarrow}

\def\be{\begin{equation}}
\def\ee{\end{equation}}
\def\g{\gamma}

\def\mt{m_t}
\def\mb{m_b}
\def\mc{m_c}

\def\bea{\begin{eqnarray}}
\def\eea{\end{eqnarray}}
\def\be{\begin{equation}}
\def\ee{\end{equation}}

\def\g{\gamma}

\def\mt{m_t}

\newcommand{\bgamaxs}{$B \to X _{s} + \gamma$}

\newcommand{\BGAMAXS}{B \ra X _{s} + \gamma}

\newcommand{\BBGAMAXS}{{\cal B}(B \ra  X _{s} + \gamma)}
\newcommand{\BBGAMAXD}{{\cal B}(B \ra  X _{d} + \gamma)}

\newcommand{\BGAMAKSTAR}{B \ra  K^{\star} + \gamma}

\newcommand{\GGAMAXS}{\Gamma (B \ra  X _{s} + \gamma)}

\def\beq{\begin{equation}}
\def\eeq{\end{equation}}

\def\Vcbabs{\vert V_{cb} \vert}

\def\Vubabs{\vert V_{ub}\vert}

\def\Vtdabs{\vert V_{td} \vert}

\def\Vtsabs{\vert V_{ts} \vert}

\newcommand{\absvcb}{\vert V_{cb}\vert}

\newcommand{\absvts}{\vert V_{ts}\vert}
\newcommand{\absvtb}{\vert V_{tb}\vert}

\newcommand{\go}[1]{\gamma^{#1}}
\newcommand{\gu}[1]{\gamma_{#1}}

\def\qbar{\overline q}

\def\q5q{\qbar{{\lambda_a}\over 2} i\gamma_5 q}

\newcommand{\bgamaxd}{$B \to X _{d} + \gamma$}

\def\to{\rightarrow}

\def\mb{m_b}

\def\as{\alpha _s}

\newcommand{\xg}{x}

\begin{document}
\begin{flushright}
DESY 97-192\\
hep-ph/9709507\\
September 1997\\
\end{flushright}
\begin{center}
{\Large \bf
\centerline{THEORY OF RARE $B$ DECAYS}}
\vspace*{1.5cm}
 {\large A.~Ali}
\vskip0.2cm
 Deutsches Elektronen Synchrotron DESY, Hamburg \\
Notkestra\ss e 85, D-22603 Hamburg, FRG\\

\vspace*{8.0cm} 
To be published in the Proceedings of the Seventh
International Symposium\\ on
Heavy Flavor Physics, University of California Santa Barbara, California,
July 7-11, 1997 %

\end{center}
\thispagestyle{empty}
\newpage   
\setcounter{page}{1}
\title{THEORY OF RARE B DECAYS}

\author{ A. Ali}

\address{Deutsches~Elektronen-Synchrotron~DESY,
     Notkestrasse~85,~D-22603~Hamburg,~FRG}

%


\maketitle\abstracts{
We discuss some selected topics in  rare $B$ decays in the
context of the standard model and compare theoretical estimates with 
available data.
Salient features of the perturbative-QCD and power corrections in the decay 
rate for $\BGAMAXS$ are reviewed and this framework is used to determine
the Cabibbo-Kobayashi-Maskawa (CKM) matrix element $\absvts$, yielding 
$\absvts =0.033 \pm 
0.007$ from the present measurements of the electromagnetic penguins.
We review estimates of the ratio $R_{K^*} \equiv
\Gamma(B \to K^* + \gamma)/\Gamma(B \to X_s + \gamma)$ 
in a number of theoretical models, which give a consistent account of this
quantity. Issues bearing on the photon energy spectrum in $\BGAMAXS$ are
also discussed. 
The CKM-suppressed decays
$B \to X_d + \gamma$, $B^\pm \to \rho^\pm + \gamma$, and
$B^{0} \to (\rho^{0}, \omega) + \gamma$ are reviewed with particular 
emphasis on the long-distance contributions in the exclusive decays. The 
impending interest in these decays in 
determining the parameters of the CKM matrix is
emphasized. Finally, the semileptonic decays $B \to X_s \ell^+ \ell^-$
are also discussed in the context of the SM.}
%
%
\section{$\BBGAMAXS$ and $\BGAMAKSTAR$ in the Standard Model and 
experiment and  determination of $\Vtsabs$} \par
\subsection{Experimental status}
Electromagnetic penguins were first sighted on the $B$ territory in 1993 
by the CLEO collaboration through the exclusive
decay $\BGAMAKSTAR$ \cite{CLEOrare1}.
This feat was followed by the daunting measurement of the inclusive 
decay $\BGAMAXS$ in 1994 by the same collaboration \cite{CLEOrare2}.
The present CLEO measurements can be summarized as \cite{Tomasz97}: 
\begin{eqnarray}  
\label{penguinexp}
{\cal B}(\BGAMAXS) &=& (2.32\pm 0.57\pm 0.35)\times 10^{-4}, \nonumber\\
{\cal B}(\BGAMAKSTAR) &=& (4.2\pm 0.8 \pm 0.6)\times 10^{-5},
\end{eqnarray}
which yield an exclusive-to-inclusive ratio:
\begin{equation}
R_{K^*} \equiv \frac{\Gamma(\BGAMAKSTAR)}{\Gamma(\BGAMAXS)}=(18.1\pm
6.8)\% ~.
 \end{equation}
Very recently, the
inclusive radiative decay has also been reported by the ALEPH collaboration 
with a (preliminary) branching ratio \cite{ALEPHbsg}: 
\begin{equation}
\label{alephbsg}
{\cal B}(H_b \to X_s + \gamma) = (3.29 \pm 0.71 \pm 0.68)\times 10^{-4}. 
\end{equation}
Since the ALEPH measurement is done at the $Z^0$ peak in the process $Z^0 
\to b \bar{b} \to H_b + X \to (X_s + \gamma) +X$,
 the  branching ratio in (\ref{alephbsg}) involves a 
different weighted average of the various $B$-mesons and $\Lambda_b$ baryons
produced in $Z^0$ decays (hence the symbol $H_b$ ) than the 
corresponding one
given in (\ref{penguinexp}), which has been measured in the decay 
$\Upsilon (4S) \to B^+ B^-, B^0 \overline{B^0}$.
Theoretically, the inclusive radiative decay widths
for the various beauty hadrons are expected to be  nearly equal.
Despite this, their branching ratios are not all equal
reflecting the differences in the respective total decay rates
(equivalently lifetimes).
 
In the context of SM, the principal interest in the decay rates in 
eqs.~(\ref{penguinexp}) and (\ref{alephbsg}) lies in that they determine the 
ratio of the CKM matrix elements \cite{CKM} $\vert V_{ts}^* 
V_{tb}/V_{cb}\vert$.
Since $\absvcb$ and $\absvtb$ have been directly measured, one can combine 
these measurements to determine $\absvts$. In addition,
the quantity $R_{K^*}$ provides information on the decay form factor in 
$\BGAMAKSTAR$. We review in this section first the branching ratio
$\BBGAMAXS$ (and ${\cal B}(H_b \to X_s + \gamma)$) in the SM and then  
discuss estimates of $\absvts$ and $R_{K^*}$.

\subsection{SM estimates of ${\cal B}(\BGAMAXS)$ and ${\cal B}(H_b \to 
X_s + \gamma)$}
The leading contribution to the decay $b \to s +\gamma$ arises
at one-loop from the so-called penguin diagrams. With the help of the
unitarity of the CKM matrix,
the decay matrix element in the lowest order can be written as:
\be
\label{e2}
{\cal M }(b \to s ~+\gamma)
    = \frac{G_F}{\sqrt{2}}\,\frac{e}{2 \pi^2} \,\lambda_{t}
(F_2 (x_t)-F_2(x_c))
 q^\mu \epsilon^\nu \bar{s} \sigma_{\mu \nu} (m_bR ~+ ~m_sL)b ~.
\ee
where $G_F$ is the Fermi coupling constant, $e=\sqrt{4 \pi 
\alpha_{\mbox{em}}}$, $x_i= ~m_i^2/m_W^2; ~i=u,c,t$ are the scaled quark 
mass ratios, and 
$q_\mu$  and $\epsilon_\mu$ are, respectively, the photon four-momentum
and polarization vector.
The GIM mechanism \cite{GIM} is manifest in this amplitude and the
CKM-matrix element dependence is factorized in $\lambda_t\equiv V_{tb} 
V_{ts}^*$.
The (modified) Inami-Lim function $F_2(x_i)$ derived from the (1-loop) 
penguin diagrams is given by \cite{InamiLim}:
\be
F_{2}(x) = 
 \frac{x}{24 (x-1)^{4}}
\times  \left[6 x (3 x -2 )
\log x - (x-1) (8 x^{2} +5 x -7 ) \right].
\ee
As the inclusive decay widths of the $B$ hadrons are proportional to
$\absvcb^2$, 
 the measurement of  $\BBGAMAXS$ can be 
readily interpreted in terms of the CKM-matrix element ratio
$\lambda_t/\Vcbabs$.
For a quantitative determination, however,  QCD radiative
and power corrections have to be computed, which we discuss next.

  The appropriate framework to incorporate
QCD corrections is that of an effective theory obtained by integrating 
out the
heavy degrees of freedom, which in the present context are the top quark 
and $W^\pm$ bosons.
 The effective Hamiltonian depends on the underlying theory and 
for the SM one has (keeping operators up to dimension 6), 
\begin{equation}\label{heffbsg}
{\cal H}_{eff}(b \to s +\gamma) = - \frac{4 G_F}{\sqrt{2}} \lambda_t
        \sum_{i=1}^{8} C_i (\mu) {\cal O}_i (\mu) ,
\end{equation}
where the operator basis, the lowest order coefficients  
$C_{i}(m_W)$ and the renormalized coefficients $C_{i}(\mu)$ can be
seen elsewhere \cite{ALI96}.
The perturbative QCD corrections to the decay rate $\GGAMAXS$ consist of two 
distinct parts:
\begin{itemize}
\item Evaluation of the Wilson coefficients
$C_i(\mu)$ at the scale $\mu=O(m_b)$.

\item Evaluation of the matrix elements of the operators
${\cal O}_i$ at the scale $\mu=O(m_b)$.
\end{itemize}
 The Wilson coefficients are calculated with the help of
the renormalization group equation whose solution requires the
knowledge of the anomalous dimension matrix in a given order in $\as$
and the matching conditions, i.e., the
Wilson coefficients $C_{i}(\mu=m_W)$, calculated in the complete theory
to the commensurate order.
 The leading logarithmic (LL) anomalous dimension matrix has been 
calculated by several independent groups   
\cite{Ciuchini}.
First calculation of the next-to-leading order (NLO)  
anomalous-dimension matrix has been carried out by Chetyrkin, Misiak and 
M\"unz \cite{Misiak96}. The matching conditions 
to order $\alpha_s$ have also been worked out in the meanwhile by several 
groups. Of these, the first six 
corresponding to the four-quark operators have been derived by Buras et 
al.~\cite{Buraswc}, and the remaining two $C_7(\mu=m_W)$ and 
$C_8(\mu=m_W)$ were worked out by Adel and Yao \cite{Yao94}. These
latter have been recalculated by Greub and Hurth \cite{Greub97},
confirming the earlier result \cite{Yao94}. Recently, these matching 
conditions
\cite{Yao94,Greub97} have also been confirmed by Buras, Kwiatkowski and Pott 
\cite{BKP97}.

  The NLO corrections to the matrix elements are of two kinds: 
 \begin{itemize}
 \item QCD Bremsstrahlung corrections $b \to s \gamma + g$, which are
needed both to cancel the infrared divergences
 in the decay rate for
$\BGAMAXS$ and in obtaining a non-trivial QCD 
contribution 
to the photon energy spectrum in the inclusive decay $\BGAMAXS$.
\item Next-to-leading order virtual corrections to the matrix elements
in the decay $b \to s +\gamma$. 
\end{itemize}
The Bremsstrahlung corrections were calculated by Ali and Greub 
\cite{ag1,ag2} in the truncated basis (involving the operators 
$O_1,~O_2,~O_7$ and $O_8$) and 
subsequently in the complete operator basis by the same 
authors \cite{ag95} and by Pott \cite{Pott95}.
The NLO virtual corrections were completed by Greub, Hurth and Wyler 
\cite{GHW96}. These latter calculations have played a key role in reducing 
the scale-dependence of the LL inclusive decay 
width. All of these pieces have been combined to get the NLO decay width
$\GGAMAXS$ and the details are given in the literature \cite{Misiak96,BKP97}.

 It is customary to express the
 branching ratio  $\BBGAMAXS$ in terms of the
semileptonic decay branching ratio ${\cal B} (B \to X\ell \nu_\ell)$,
\begin{equation}
\label{brbsgsm}
{\cal B} ( B \ra  X_{s} \g) = [\frac{\Gamma(B \ra  
\gamma + X_{s})}{\Gamma_{SL}}]^{th}
\, {\cal B} (B \to X\ell \nu_\ell), 
\end{equation}  
and the theoretical part can be expressed as \cite{Misiak96} 
\begin{equation}
\label{brdef}
[\frac{\Gamma(B \ra\gamma + X_{s})}{\Gamma_{SL}}]^{th}=
\frac{ \vert\lambda_t\vert^2}{V_{cb}^2} \frac{6 \alpha}{\pi f(z)}F(\vert D 
\vert^2 +A)~.
 \end{equation}
Here, 
\begin{eqnarray}
f(z) &=& 1-8z + 8z^3 -z^4 -12z^2\ln z ~~~\mbox{with} ~~~z=\frac{m_{c,pole}^2}
{m_{b,pole}^2} \nonumber\\
F &=& \frac{1}{\kappa(z,\bar{\mu}_b)} \left( 
\frac{\overline{m_b}(\mu=m_b)}{m_{b,pole}}\right)^2
= \frac{1}{\kappa(z,\bar{\mu}_b)}\left(1-\frac{8}{3} 
\frac{\alpha_s(m_b)}{\pi}\right)~,
 \end{eqnarray}
where $m_{b,pole}$ ($m_{c,pole}$) is the $b$ ($c$)-quark pole mass and
$\overline{m_b}(\mu=m_b)$ is the $b$-quark mass in the 
$\overline{\mbox{MS}}$ scheme. The function
$\kappa(z,\bar{\mu}_b)$ represents the QCD correction to the semileptonic 
decay 
width \cite{CM78}, which depends on the scale $\bar{\mu}_b =O(m_b)$; 
its analytic form is given by Nir \cite{Nir89}. The other functions ($D$ and 
$A$) can be seen in literature \cite{Misiak96}. It should be remarked that 
while the
Bremsstrahlung function $A$ has been calculated in the complete operator 
basis, the
virtual corrections contributing to $D$ are still known in the truncated
approximation.
 However, since the numerical values of the Wilson 
coefficients $C_3,...,C_6$ are very small, the contribution left out in the 
NLO expression in eq.~(\ref{brdef}) is expected to be 
small (not more than several percent).

In addition to the perturbative QCD improvements discussed above, also the
leading power corrections, which start in $1/\mb^2$, have been 
calculated to the 
decay widths appearing in the numerator and denominator of eq.~(\ref{brdef}) 
\cite{georgi,manoharwise,FLS94}.
The power corrections in the numerator have been obtained assuming that the
decay $\BGAMAXS$ is dominated by the magnetic moment operator $O_7$.
Writing this correction in an obvious notation as
\begin{equation}
\frac{\Gamma(\BGAMAXS)}{\Gamma^{0}(\BGAMAXS)} = 1 + \frac{\delta_b}{m_b^2},
\end{equation}
one obtains $\delta_b =1/2 \lambda_1 -9/2 \lambda_2$, where $\lambda_1$ and 
$\lambda_2$ are, respectively, the kinetic energy and magnetic moment 
parameters of the theoretical framework based on heavy quark expansion
(henceforth called HQET) \cite{Mannel97}. Using $\lambda_1 =-0.5 
~\mbox{GeV}^2$ and $\lambda_2 
=0.12 ~\mbox{GeV}^2$, one gets $\delta_b/\mb^2 \simeq -4\%$. However, it 
turns out
 that the leading order $(1/m_b^2)$ power 
corrections in the heavy quark expansion are identical 
in the inclusive decay rates  $\Gamma(\BGAMAXS)$ and $\Gamma(B \to X \ell 
\nu_\ell)$,
as far as $\lambda_1$ is concerned.  The corrections proportional
$\lambda_2$ differ only marginally. Thus, including or
neglecting the $1/m_b^2$ corrections makes a difference of only $1\%$ in the
ratio (\ref{brdef}) and hence in $\BBGAMAXS$.

Recently, the power corrections proportional to $1/m_c^2$, resulting from the
interference of the operators $O_2$ and $O_7$ in $\BGAMAXS$, have also been
worked out \cite{Voloshinbsg,powermc,BIR97}. 
Expressing this symbolically as 
\begin{equation}
\frac{\Gamma(\BGAMAXS)}{\Gamma^{0}(\BGAMAXS)} = 1 + \frac{\delta_c}{m_c^2},
\end{equation} 
one finds \cite{BIR97} $\delta_c/m_c^2 \simeq + 0.03$.

There exist several (marginally) differing numerical values of the
branching ratio ${\cal B}(\BGAMAXS)$ in the SM. 
Using $|V_{ts}^* V_{tb}/V_{cb}|=0.976 \pm 0.010$ obtained from the unitarity 
constraints \cite{PDG96}, fixing the scale $\bar{\mu}_b=m_b$, varying the 
scale $\mu_b$ in 
the range $2 m_b \geq \mu_b \geq ~m_b/2$, and using current values of 
the various input parameters, but ignoring the $\delta_c/m_c^2$ term,
the short-distance contribution  
has been estimated by Chetyrkin et al.~as 
\cite{Misiak96}: ${\cal B} (\BGAMAXS )= (3.28 \pm 0.33) \times 10^{-4}$,
where all the errors have been combined in quadrature. Including the
$\delta_c/m_c^2$ term, setting $\bar{\mu}_b=\mu_b$
and varying $\mu_b$ in the stated range, but keeping the other
parameters the same as in the work of Chetryrkin et al. \cite{Misiak96}, 
Greub and Hurth \cite{GH97b} determine ${\cal B} (\BGAMAXS )= (3.38 \pm 
0.33) \times 10^{-4}$. Recently, a NLO value 
has been obtained by Buras et al. \cite{BKP97}, yielding
${\cal B} (\BGAMAXS )= (3.48 \pm 0.31) \times 10^{-4}$.
The shift in the central value from the one given by 
Greub and Hurth \cite{GH97b}
is due to systematically discarding the next-next-leading order terms, which
were kept in the earlier work \cite{Misiak96,GH97b}, and the reduced 
error is due to treating the scale uncertainty in the numerator and 
denominator in eq.~(\ref{brdef}) independently.   

 The numerical values for ${\cal B} ( B \ra  X_{s} \g)$ 
quoted above
\cite{Misiak96,BKP97,GH97b} have been obtained by
using the semileptonic branching ratio ${\cal B}(sl) =(10.4 \pm 0.4)\%$ 
taken from earlier data at $\Upsilon(4S)$.
This number has changed somewhat in the meanwhile and
the current measurements are  
\cite{Feindt97}:  ${\cal B}(sl) =(10.49 \pm 0.46)\%$
(at $\Upsilon(4S))$ versus ${\cal B}(sl) =(11.16 \pm 0.20)\%$ (at $Z^0$).
Updating ${\cal B}(sl)$  yields
\begin{equation}\label{upssmali}
{\cal B} (\BGAMAXS )= (3.51 \pm 0.32) \times 10^{-4}~,
\end{equation}
to be compared with the CLEO measurement ${\cal B} (\BGAMAXS )= (2.32 \pm 
0.67) \times 10^{-4}$. The corresponding inclusive branching ratio at the 
$Z^0$ is obtained by using in eq.~(\ref{brbsgsm}) the semileptonic
branching ratio measured at the $Z^0$. This gives
\begin{equation}\label{zsmali}
{\cal B} (H_b \to X_s + \gamma )= (3.76 \pm 0.30) \times 10^{-4}~,
\end{equation}
to be compared with the ALEPH measurement
${\cal B} (H_b \to X_s + \gamma )= (3.29 \pm
0.98) \times 10^{-4}$.  
The agreement between experiment and SM is good though the CLEO number is
marginally ($2\sigma)$ lower than the SM branching ratio.

We would like to use the NLO SM-based theory and experiments to determine 
the CKM matrix element ratio $|V_{ts}^* V_{tb}/V_{cb}|$ and
$|V_{ts}|$. The two inclusive measurements given in eqs.~(\ref{penguinexp}) 
and (\ref{alephbsg}), and the corresponding NLO SM-estimates given in
eqs.~(\ref{upssmali}) and (\ref{zsmali})  yield,
 \begin{eqnarray}
\label{vtscb}
|\frac{V_{ts}^* V_{tb}}{V_{cb}}| &=& 0.79 \pm 0.11 ~(\mbox{expt}) \pm 0.04
~(\mbox{th}) ~(@ 
\Upsilon(4s)), \nonumber\\
|\frac{V_{ts}^* V_{tb}}{V_{cb}}| &=& 0.91 \pm 0.14  ~(\mbox{expt})
\pm 0.04 ~(\mbox{th}) ~(@ Z^0). 
 \end{eqnarray}
Averaging the two measurements following the PDG prescription \cite{PDG96}
gives the following weighted average for the CKM matrix element ratio: 
\begin{eqnarray}
|\frac{V_{ts}^* V_{tb}}{V_{cb}}| &=& 0.84  \pm  0.09 ~(\mbox{expt}) \pm 
0.04 ~(\mbox{th}) \nonumber\\
&\Longrightarrow & 0.84  \pm  0.10
\label{vcs}
\end{eqnarray}
where the second row has been obtained  by
adding the theoretical and experimental errors in quadrature. With the CKM
unitarity, one has $ |\frac{V_{ts}^* V_{tb}}{V_{cb}}| \simeq |V_{cs}|$;
this equality holds numerically (within present precision) if one compares 
the l.h.s.
obtained from the decay $B \to X_s + \gamma$  given in eq.~(\ref{vcs}) 
with the present determination of the r.h.s. from charmed hadron decays 
\cite{PDG96}, $|V_{cs}|= 1.01 \pm 0.18$.
Using the value of $\absvtb$
measured  by the CDF collaboration \cite{CDFvtb}, $\vert V_{tb} 
\vert = 0.99 \pm 0.15$ and noting \cite{Gibbons96} that $\absvcb = 0.0393 
\pm 0.0028$, finally yields
 \begin{equation}
\absvts = 0.033 \pm 0.007~,
\end{equation}
where all the errors have been added in quadrature.
This is probably as direct a determination of $\vert V_{ts}\vert$ as we will 
ever see, as the  decay $t \to W +s$ is too daunting to measure due
to the low tagging efficiency of the $s$-quark jet. With improved measurement
of ${\cal B}(\BGAMAXS)$ and $V_{tb}$, one expects to reduce the 
present error on $\absvts$ by a factor of 2, possibly 3.

The exclusive-to-inclusive ratio $R_{K^*}$ has been worked out in a
number of models. This involves estimates of 
the matrix elements of the electromagnetic penguin operator,
implicitly assuming the SD-dominance.
Taken on their face value, different models give a rather 
large theoretical dispersion on $R_{K^*}$. However, one should stress that 
QCD sum rules, models based on quark-hadron duality, and the improved
lattice-QCD estimates of the 1997 vintage (being discussed by Lynn 
\cite{Flynn97}) are theoretically more reliable. Concentrating only on them,
some representative results are:\\
$R_{K^*} = 0.20 \pm 0.06 ~~[\mbox{Ball ~\protect\cite{bksnsr}}],
~R_{K^*} = 0.17 \pm 0.05 ~~\left[\mbox{Colangelo et al.
~\protect\cite{bksnsr}}\right],\\
~R_{K^*} = 0.16 \pm 0.05 ~~\left[\mbox{Ali, Braun \& Simma
~\protect\cite{abs93}}\right], 
~R_{K^*} = 0.16 \pm 0.05 ~~\left[\mbox{Narison
~\protect\cite{bksnsr}}\right], \\
~R_{K^*} = 0.13 \pm 0.03 ~~\left[\mbox{Ali \& 
Greub~\protect\cite{ag1}}\right], 
~R_{K^*} = 0.16 ^{+ 0.04}_{-0.03} ~~\left[\mbox{Flynn 
~\protect\cite{Flynn97}}\right]$.
These estimates are consistent with each other and with the CLEO
measurement $R_{K^*}=0.181 \pm 0.06$.
Summarizing this section, it is fair to conclude that SM gives a 
quantitative account of data in electromagnetic penguin decays
in inclusive rates, yielding a first determination of $\absvts$ 
with an accuracy of  $\pm 20\%$. Further, both $\BBGAMAXS$ and the ratio 
$R_{K^*}$ are in agreement with the dominance of the short-distance physics
in these decays.
%
%
%
\section{Photon energy spectrum in $\BGAMAXS$}
Calculation of the photon energy spectrum is  somewhat
intractable as the main work-horse, namely HQET, does not quite make it to 
the very end of this spectrum. There is no alternative at present but to
model the non-perturbative effects. We review the present state of the art.

The two-body partonic process $b \to s \gamma$
yields a photon energy spectrum which is just a discrete line, $1/(\Gamma) d 
\Gamma (b \to s \gamma) = \delta(1-x)$, where 
the scaled photon energy $x$ is defined as
$ x\equiv 2\, E_\g \mb /(\mb^2-m_s^2 )$. The physical photon energy 
spectrum is obtained by convoluting the non-perturbative effects and the
perturbative QCD corrections, such as the ones arising from the decay
 $b \to s \gamma + g$. The latter gives a
characteristic Bremsstrahlung spectrum in $x$ in the interval $[0,1]$
peaking near the end-points, $E_\gamma \to
E_\gamma ^{max}$ (or $x \to 1$) and $E_\gamma \to 0$ (or $x \to 0$), arising 
from the soft-gluon and soft-photon configurations, respectively.
Near the end-points, one has to improve the spectrum obtained in fixed order
perturbation theory.
This is  done in the region $x \to 1$ by isolating and 
exponentiating the leading behaviour in $\alpha_{em}\alpha_s(\mu)^m 
\log^{2n} (1-x)$  with $m\leq n$,
where $\mu$ is a typical momentum in the decay $\BGAMAXS$. In this region,
which is dominated
by the magnetic moment operator $O_7$, the spectrum  can be symbolically
 expressed as 
 \begin{equation}
\frac{d\Gamma_{77}^{\exp}}{d\xg} = -C \frac{\alpha_s(\mu)}{3 
\pi}(C_7^{eff})^2 
\exp\left(\frac{\alpha_s(\mu)}{3 \pi} \Omega_1\right) \left[ 
\Omega_2^\prime + 
\Omega_1^\prime \left( 1+ \frac{\alpha_s(\mu)}{3 \pi} \Omega_2 
\right)\right]\, ,
 \end{equation}
where $C$ is a normalization constant ($r\equiv m_s/m_b$).
\begin{equation}
C= (1-r)^3(1+r) \frac{\overline{m_b}(\mu=\mb)^2 \mb(pole)^3}{32 \pi^4}
\alpha G_F^2 |\lambda_t|^2
\end{equation}

 The running of
$\alpha_s$ is a non-leading effect, but  as it is characteristic of QCD it
modifies the Sudakov-improved end-point photon energy spectrum
\cite{shifmangamma} compared to its  analogue in QED \cite{Sudakov}.
 The expressions for $\Omega_1 (\xg,r)$, 
$\Omega_2(\xg,r)$ and their derivatives (denoted by primes) 
can be seen for non-zero $s$-quark mass in literature\cite{ag1,ag2,ag95}. 
The expressions (with $r \to 0$) are:
\begin{eqnarray}
\Omega_1 &=& -2 \ln^2(1-\xg) - 7 \ln (1-\xg), \nonumber\\
\Omega_2 &=& 10(1-\xg) + (1-\xg)^2 -\frac{2}{3} (1-\xg)^3 -(1-\xg) (3+\xg) 
\ln(1-\xg)\,,
\end{eqnarray}
where the double logarithmic term in $\Omega_1$ is universal \cite{Sudakov}. 
The other terms are specific to the decay $B \to X_s + \gamma$. 
 
 As long as the $s$-quark mass
is non-zero, there is no collinear singularity in the spectrum.
However, parts of the spectrum have large logarithms of the form
$\as \log (m_b^2/m_s^2)$, which are important near the end-point
$\xg \to 0$ and are, in principle, present for any photon energy and 
should be
resummed. The order $\alpha_s$ corrected distribution
for low photon energies is dominated by the operator $O_8$ and is 
given by\cite{ag95,klp95}:
\begin{equation}  
\frac{d\Gamma_{88}}{d\xg}|_{\xg >0} = C \frac{\alpha_s(\mu)}{3 \pi} 
 \frac{(C_8^{eff})^2}{9} \tilde{\Gamma}(x,r)\, ,
\end{equation}
where \begin{eqnarray}
\tilde{\Gamma}(x,r) &=&
\left(\frac{4+4r}{\xg-r\xg} -4 + 2\xg\right) 
 \ln \frac{1-\xg +r\xg}{r} \nonumber\\
&-&\frac{(1-x)\left[8-(1-r)\xg(16-9\xg+7r\xg) + (1-r)^2 \xg^3 (1-2\xg) 
\right]} {\xg(1-\xg+r\xg)^2} .
\end{eqnarray}
The difference between the fixed order spectrum and its exponentiated 
version 
can be seen in the work of Kapustin et al.~\cite{klp95}.
 If low energy photons can be
detected in $B \to X_s + \gamma$ (say, for $E_\gamma \leq 1 $ GeV), then 
such an experiment could help measure $C_8^{eff}$. Unfortunately, with 
the SM values 
of the Wilson coefficients, the partial branching ratio for $\BGAMAXS$
for low photon energies is too small to be measured even at $B$ factories.
However, with an anomalously large $C_8^{eff}$, as has been entertained
in the literature in other contexts \cite{kagan95,cgg96}, the
$C_8^{eff}$-dependent part of the photon energy spectrum may get appreciably
enhanced. It is worthwhile to measure the spectrum in the 
intermediate energies ($ 1.0 ~\mbox{GeV} \leq 2.0 ~\mbox{GeV}$) to search 
for the effect of such anomalously enhanced $C_8^{eff}$ contribution in
beyond-the-SM scenarios.

\par
 Implementation of non-perturbative effects is at present a
model dependent enterprise and data are not precise enough 
to distinguish various models proposed 
in the literature \cite{ag95,shifmangamma,LY96}. We shall confine ourselves 
to the discussion of the photon energy spectrum calculated in a simple
model, in which the $b$ quark in the $B$ hadron is assumed to have a Gaussian
distributed Fermi motion \cite{Aliqcd}
determined by a non-perturbative parameter, $p_F$.
This model describes well the lepton energy 
spectrum in semileptonic decays $B \to X \ell \nu_\ell$ and
it has also received some theoretical support in the HQET approach 
subsequently.

The photon energy spectrum based on this model, including the 
QCD perturbative improvements, 
has been used both by the CLEO \cite{CLEOrare2} and ALEPH collaboration
\cite{ALEPHbsg} in the analysis of their data on $\BGAMAXS$.
An analysis of the CLEO photon 
energy spectrum has also been undertaken \cite{ag95} to determine 
the non-perturbative parameters of this model, namely  
$m_b(pole)$ and $p_F$.  The latter is related to the
kinetic energy parameter $\lambda_1$ defined earlier in the HQET approach. 
The minimum $\chi^2$ of the CLEO data is obtained for $p_F=450$ MeV and 
$m_b(pole)=4.77$ GeV. However, the $\pm 1 \sigma$ errors on these quantities
are large (a similar conclusion has been drawn in terms of $\lambda_1$ 
and $m_b(pole)$ by Li and Yu \cite{LY96}). The interesting question here is
to determine if the non-perturbative aspects in the decays $B \to X \ell 
\nu_\ell$ and $\BGAMAXS$ can be described in terms of a universal shape 
function.
To pursue this further requires lot more data which we hope will soon be 
forthcoming.
%
%
\section{Inclusive radiative decay \bgamaxd\ and constraints on the CKM 
parameters} %
\par
The quantity of interest in the
decay $B \to X_d + \gamma$ is the high energy part of the photon energy 
spectrum, which has to be measured requiring that
 the hadronic system $X_d$ recoiling against the
photon does not contain strange hadrons to suppress the large-$E_\g$
photons from the decay $\BGAMAXS$, which is now the largest background.
 Assuming that  such an experiment is feasible,
one can determine  from the ratio of the branching ratios
$\BBGAMAXD/\BBGAMAXS$ the parameters of the CKM
matrix (in particular $\rho$ and $\eta$ in the Wolfenstein 
parameterization \cite{Wolfenstein}).

\indent
 In close analogy
with the \bgamaxs\ case discussed earlier,
the complete set of dimension-6 operators relevant for
the processes $b \to d \gamma$ and $b \to d \gamma g$ 
can be written as:
\begin{equation}
\label{heffd}
{\cal H}_{eff}(b \to d)=
 - \frac{4 G_{F}}{\sqrt{2}} \, \xi_{t} \, \sum_{j=1}^{8}
C_{j}(\mu) \, \hat{O}_{j}(\mu),\quad
\end{equation}
where $\xi_{j} = V_{jb} \, V_{jd}^{*}$ with $j=u,c,t$. The operators
 $\hat{O}_j, ~j=1,2$, have implicit in them CKM factors.
We shall use the Wolfenstein parametrization \cite{Wolfenstein},   
in which case the matrix is determined in terms of the four parameters
$A, \lambda=\sin \theta_C$, $\rho$ and $\eta$, and one can express the above
factors as :
\begin{equation} 
\xi_u = A \, \lambda^3 \, (\rho - i \eta),
~~~\xi_c = - A \, \lambda^3 ,
~~~\xi_t=-\xi_u - \xi_c.
\end{equation}
We note that all three CKM-angle-dependent quantities
$\xi_j$ are of the
same order of magnitude, $O(\lambda^3)$. It is convenient to define 
the operators $\hat{O}_1$ and 
$\hat{O}_2$ entering in ${\cal H}_{eff}(b \to d)$ as follows \cite{ag2}:
\begin{eqnarray}
\label{basis}
&&\hat{O}_{1} =
 -\frac{\xi_c}{\xi_t}(\bar{c}_{L \beta} \go{\mu} b_{L \alpha})
(\bar{d}_{L \alpha} \gu{\mu} c_{L \beta}
 -\frac{\xi_u}{\xi_t}(\bar{u}_{L \beta} \go{\mu} b_{L \alpha})
(\bar{d}_{L \alpha} \gu{\mu} u_{L \beta}) ,\nonumber \\
&& \hat{O}_{2} =
-\frac{\xi_c}{\xi_t}(\bar{c}_{L \alpha} \go{\mu} b_{L \alpha})
(\bar{d}_{L \beta} \gu{\mu} c_{L \beta}) 
 -\frac{\xi_u}{\xi_t}(\bar{u}_{L \alpha} \go{\mu}
b_{L \alpha}) (\bar{d}_{L \beta} \gu{\mu} u_{L \beta}) ,
\end{eqnarray}
with the rest of the operators $(\hat{O}_j;~j=3...8)$ 
defined like their
counterparts ${O}_j$ in ${\cal H}_{eff}(b \to s)$, with the obvious 
replacement
$s \to d$. With this choice, the matching conditions $C_j(m_W)$
 and the solutions
of the RG equations yielding $C_j(\mu)$ become
identical for the two operator bases $O_j$ and $\hat{O}_j$.
The branching ratio  $\BBGAMAXD$ in the SM  can be generally written as:
\bea
\label{branstruc}
\lefteqn{\BBGAMAXD = D_1 \lambda^2}
\nonumber\\&&{}
 \{(1-\rho)^2 + \eta^2 -(1-\rho) D_2 - \eta D_3 +D_4  \} , \quad
\eea
where the functions $D_i$ depend on various parameters such 
as $\mt,\mb,\mc,\mu$, and $\as$.
These functions were calculated in the LL approximation some time ago 
\cite{ag2}  and since then their estimates have been improved 
\cite{aag96}, making use of the NLO calculations discussed earlier. We shall
assume, based on model calculations  \cite{DHT95,GP95,Ricciardi}
and the $1/m_c^2$ power corrections discussed earlier in the context of
$\BGAMAXS$, that the LD contributions are small also in $\BBGAMAXD$.

 To get an estimate of  
$\BBGAMAXD$ at present, the CKM parameters $\rho$ and $\eta$ have 
to be constrained from the unitarity fits, which yield the 
following ranges (at 95\% C.L.) \cite{al96}:
\begin{eqnarray}
 0.20 &\leq & \eta \leq 0.52 , \nonumber \\
 -0.35 &\leq & \rho \leq 0.35 ~.
\label{rhoetarange}
\end{eqnarray}
The (nominally) preferred CKM-fit values at present are
$(\rho,\eta) = (0.05,0.36)$,
for which one gets \cite{aag96} 
\begin{equation}
\BBGAMAXD = (1.63 \pm 0.16) \times 10^{-5},
\end{equation}
where the error estimate follows from the one for $\BBGAMAXS$.
Allowing the CKM parameters to vary over the entire allowed domain, one 
gets (at 95\% C.L.) 
\begin{equation}
 6.0 \times 10^{-6} \leq \BBGAMAXD \leq 3.0 \times 10^{-5}.
\end{equation} 
The present theoretical uncertainty in this rate is a factor 5,
 which shows that even a modest 
measurement of $\BBGAMAXD$ will have a very significant impact on the
CKM phenomenology. To the best of our knowledge, there is no experimental
bound available on $\BBGAMAXD$, but we hope that this decay will be 
measured in future at the $B$ factories and CLEO.
\section{CKM-suppressed exclusive decays ${\cal B}(B \to V + \gamma )$ } 
\par
Exclusive radiative
 $B$ decays $B \to V + \gamma$, with $V=K^*,\rho,\omega$, are also 
potentially very interesting for the CKM phenomenology
\cite{abs93}. Extraction of CKM parameters would, however, involve a 
trustworthy estimate of the SD- and LD-contributions in the decay amplitudes.
We have argued that the decays $\BGAMAXS$ and $(B^\pm,B^0) \to 
(K^{*\pm},K^{*,0}) + \gamma$ are consistent with the dominance of the
SD-contribution. 
There exist good reasons to believe that also the CKM-suppressed exclusive 
radiative decays are dominated by SD-physics, though one has to work out
the LD-contribution on a case-by-case basis. 
 More importantly, data on the various charged and neutral $B$ meson 
radiative decays can be 
used directly to put meaningful bounds on the LD-contributions. 
Hence, despite skepticism in some quarters and in this conference 
\cite{buras97}, we 
believe that exclusive radiative $B$ decays are worth measuring.
 
 \par
  The SD-contribution in the 
 exclusive decays $(B^\pm, B^{0}) \to (K^{*\pm}, K^{* 0})+ \gamma$,
$(B^\pm, B^{0}) \to (\rho^\pm,\rho^{0}) + \gamma$,
$B^{0} \to \omega + \gamma$  and the
corresponding $B_s$ decays, $B_s \to \phi + \gamma $, and
$B_s \to K^{* 0} + \gamma $,
involve the magnetic moment operator ${\cal O}_7$ and the related one 
obtained by the obvious change $s \to d$, $\hat{O}_7$.
The transition form factors governing the radiative $B$ decays
 $B \to V + \gamma$ can be generically  defined as:
\be
 \langle V,\lambda |\frac{1}{2} \bar \psi \sigma_{\mu\nu} q^\nu b
 |B\rangle  =
     i \epsilon_{\mu\nu\rho\sigma} e^{(\lambda)}_\nu p^\rho_B p^\sigma_V
F_S^{B\rightarrow V}(0).
\label{defF}
\ee
Here $V$ is a vector meson
with the polarization vector $e^{(\lambda)}$ and $\psi$ stands for the
field of a light $u,d$ or $s$ quark.
 In (\ref{defF}) the QCD
renormalization of the $\bar \psi \sigma_{\mu\nu} q^\nu b$ operator
is implied. Keeping only the SD-contribution 
 leads to obvious relations among the exclusive 
decay rates, exemplified here by the decay
rates for $(B^\pm,B^0) \to \rho + \gamma$ and $(B^\pm,B^0) \to K^* + 
\gamma$: 
\be
\frac{\Gamma ((B^\pm,B^{0}) \to (\rho^\pm,\rho^{0}) + \gamma)}
     {\Gamma ((B^\pm,B^{0}) \to (K^{*\pm},K^{* 0}) + \gamma)} 
  \simeq \kappa_{u,d}\left[\frac{\Vtdabs}{\Vtsabs}\right]^2 \,,
\label{SMKR}
\ee
where $\kappa_{i} \equiv [F_S(B_i \to \rho \gamma)/F_S(B_i \to K^* 
\gamma)]^2$, which is unity in the $SU(3)$ limit. (This is not being
recommended as the $SU(3)$-breaking effects have been calculated in a
number of papers \cite{bksnsr,abs93}.)
 Likewise, assuming 
dominance of SD physics gives relations among various decay rates
 \beq\label{ratio2}
\Gamma(B^\pm \to \rho^\pm \gamma)=2 ~\Gamma(B^{0}\to \rho^0  \gamma)
    = 2 ~\Gamma (B^{0} \to \omega  \gamma)~,
\eeq
where the first equality holds due to the isospin invariance,
and in the second $SU(3)$ symmetry has been assumed.

The LD-amplitudes in radiative $B$ decays from the light quark 
intermediate states necessarily involve other CKM matrix elements. 
In the CKM-suppressed decays $B \to V + \gamma$ they  are
dominantly induced by the matrix elements of the
four-Fermion operators $\hat{O}_1$ and $\hat{O}_2$. 
 Estimates of these contributions have been obtained
in the light-cone QCD sum rule approach \cite{wyler95,ab95}
Using factorization, the LD-amplitude in the decay $B^\pm \to \rho^\pm + 
\gamma$ can be written in terms of the form factors $F_1^L$ and $F_2^L$,
\begin{eqnarray}\label{Along}
{\cal A}_{long} &=&
-\frac{e\,G_F}{\sqrt{2}} V_{ub}V_{ud}^\ast
\left( C_2+\frac{1}{N_c}C_1\right) m_\rho
\varepsilon^{(\gamma)}_\mu \varepsilon^{(\rho)}_\nu
\nonumber\\&&{}\times
 \Big\{-i\Big[g^{\mu\nu}(q\cdot p)- p^\mu q^\nu\Big] \cdot 2 F_1^{L}(q^2)
 +\epsilon^{\mu\nu\alpha\beta} p_\alpha q_\beta
 \cdot 2 F_2^{L}(q^2)\Big\}\,.
\end{eqnarray}
The two form factors are found to be  numerically close to each other in the 
QCD sum rule approach, $F_1^L\simeq F^L_2 \equiv F_L$,
hence the ratio of the LD- and the SD- contributions reduces to a number 
\cite{ab95}
 \begin{equation}\label{ratio2p}
{\cal A}_{long}/{\cal A}_{short}=
R_{L/S}^{B^\pm\to\rho^\pm\gamma}
\cdot\frac{V_{ub}V_{ud}^\ast}{V_{tb}V_{td}^\ast} ~.
\end{equation}
where 
\be
\label{result2}
R_{L/S}^{B^\pm\to\rho^\pm\gamma}  \equiv 
 \frac{4 \pi^2 m_\rho(C_2+C_1/N_c)}{m_b C_7^{\mathit{eff}}}
\cdot\frac{F_L^{B^\pm \to \rho^\pm \gamma}}{F_S^{B^\pm \to \rho^\pm 
\gamma}} = -0.30\pm 0.07 ~.
\ee

The analogous LD-contributions to the neutral $B$ decays
$B^{0}\to\rho\gamma $ and $B^{0}\to\omega\gamma $ are
expected to be much smaller:
\begin{equation}
\frac{R_{L/S}^{B^{0}\to\rho\gamma}}{R_{L/S}^{B^\pm\to\rho^\pm\gamma}}=
\frac{e_d a_2}{e_u a_1} \simeq -0.13 \pm 0.05 ,
\end{equation}
where the numbers are based on using \cite{BH95}
$a_2/a_1 = 0.27 \pm 0.10$ and $e_d/e_u =-1/2$ is the ratio of the 
electric charges for the $d$- and $u$-quarks. This would then yield
$R_{L/S}^{B^{0}\to\rho\gamma} \simeq R_{L/S}^{B^{0}\to\omega\gamma}=0.05$.

 To get a ball-park estimate of the ratio
${\cal A}_{long}/{\cal A}_{short}$, we take the central value from 
the CKM fits, yielding  \cite{al96} $\Vubabs/\Vtdabs \simeq 0.33$,
which in turn gives,
 \bea
\label{bpmld}
|{\cal A}_{long}/{\cal A}_{short}|^{B^\pm\to\rho^\pm\gamma}
&=& |R_{L/S}^{B^\pm\to\rho^\pm\gamma}|
\frac{|V_{ub}V_{ud}|}{|V_{td}V_{tb}|}
 \simeq 0.1 ~, \nonumber\\
 \frac{{\cal A}_{long}^{B^{0}\to\rho\gamma}}{{\cal 
A}_{short}^{B^{0}\to\rho\gamma}}& \leq& 0.02.
\eea
 That the LD-effects remain small
in ${B^{0}\to\rho\gamma}$ decay has also been supported in an analysis
based on the soft-scattering of on-shell hadronic decay products
$B^{0} \to \rho^0 \rho^0 \to \rho \gamma$ \cite{DGP96},
though this paper estimates them somewhat higher   
(between $4\%$ and $8\%$).

 The relations (\ref{ratio2}), which obtain ignoring LD-contributions,
get modified by including the LD-contributions to
\begin{equation}\label{ratio5}
\frac{\Gamma(B^\pm\to \rho^\pm\gamma)}{2\Gamma(B^{0}\to \rho\gamma)}
=\frac{\Gamma(B^\pm\to \rho^\pm\gamma)}{2\Gamma(B^{0}\to \omega\gamma)}
=1+ \Delta (R_{L/S})\, ,
\end{equation}
where ($R_{L/S}\equiv R_{L/S}^{B^\pm\to\rho^\pm\gamma}$)
\begin{equation}
\Delta (R_{L/S})=2\cdot R_{L/S} 
V_{ud}\frac{\rho(1-\rho)-\eta^2}{(1-\rho)^2+\eta^2}
+(R_{L/S})^2 V_{ud}^2\frac{\rho^2+\eta^2}{(1-\rho)^2+\eta^2}\,.
\end{equation}

\par
The ratio of the CKM-suppressed and CKM-allowed
decay rates  for charged $B$ mesons
gets likewise modified due to the LD contributions. Following earlier 
discussion,
we ignore the LD-contributions in $\Gamma(B \to K^*\gamma)$. The ratio of
the decay rates in question can therefore be written as:
\begin{equation}\label{ratio3}
\frac{\Gamma(B^\pm\to \rho^\pm\gamma)}{\Gamma(B^\pm\to 
K^{*\pm}\gamma)} = \kappa_u \lambda^2[(1-\rho)^2+\eta^2]
\left(1+ \Delta (R_{L/S})\right)
\end{equation}
The effect of the LD-contributions is modest but not negligible, introducing
an uncertainty  
comparable to the $\sim 15\%$ uncertainty in the overall normalization
due to the $SU(3)$-breaking effects in the quantity $\kappa_u$.

\indent
Neutral $B$-meson radiative decays are less-prone to the LD-effects,
 as argued above, and hence one expects that to a good approximation
(say, better than $10\%$) the ratio of the decay rates for neutral $B$ meson 
obtained in the approximation of SD-dominance remains valid \cite{abs93}:
\begin{equation}
\frac{\Gamma(B^0\to \rho\gamma,\omega\gamma)}{\Gamma(B\to K^*\gamma)}
 = \kappa_d\lambda^2 [(1-\rho)^2+\eta^2]~,
\end{equation}
where this relation holds for each of the two decay modes separately.
This ratio is at par with the mass-difference ratio $\Delta M_d/\Delta M_s$
in the neutral $B$-meson sector, as both involve a reliable estimate of the
$SU(3)$-breaking effects but otherwise reflect the dominance of the
SD-physics. 

 Finally, combining the estimates for the LD- and SD-form factors
\cite{ab95} 
\cite{abs93},  and restricting the Wolfenstein
parameters in the allowed range given earlier, yields
\begin{eqnarray}\label{ratio4}
{\cal B}(B^\pm\to \rho^\pm\gamma)
&=& (1.5 \pm 1.1) \times 10^{-6} ~,
\nonumber\\{}
{\cal B}(B^{0}\to \rho\gamma) & \simeq & {\cal B}(B^{0}\to \omega \gamma)
\nonumber\\{}
& = & (0.65 \pm 0.35) \times 10^{-6} ~,
\end{eqnarray}
where we have used the experimental value for the branching ratio
${\cal B} (B \to K^* + \gamma)$
\cite{CLEOrare1}.
 The large range reflects to a large extent the poor
knowledge of the CKM matrix elements and hence experimental measurements
of these branching ratios will contribute greatly to determine the
Wolfenstein parameter $\rho$ and $\eta$.
 Present experimental limits 
(at $90\%$ C.L.) are \cite{Tomasz97}:
${\cal B}(B^\pm\to \rho^\pm\gamma) <1.1 \times 10^{-5}$,
${\cal B}(B^{0}\to \rho\gamma) < 3.9 \times 10^{-5}$ and
${\cal B}(B^{0}\to \omega \gamma) < 1.3 \times 10^{-5}$. The constraints
on the parameters $(\rho,\eta)$ following from them are, however, not yet
competitive to the one following from unitarity and lower bound on
the mass difference in the $B_s^0$ - $\overline{B_s^0}$ sector \cite{al96}. 
\section{Inclusive rare decays $B \to X_s \ell^+ \ell^-$ in the SM}
\par
The decays \bxsll, with $\ell=e,\mu,\tau$, provide new avenues to 
search for physics beyond the standard model 
\cite{Masieroetal,AGM94,CMW96,Gotoetal96,Hewett97}.
The branching ratio $\BBGAMAXS$ constrains
the magnitude of $C_7^{\mathit{eff}}$ but
 the sign of $C_7^{\mathit{eff}}$ is not
determined by the measurement of ${\cal B}(\BGAMAXS)$. This sign 
is in general model dependent.
It is known that
in SUSY models, both the negative and positive signs are 
allowed as one scans over the allowed SUSY parameter space.
The \bxsll ~amplitude in the standard model (as well as in several 
extensions of it such as SUSY) depends on the coefficient 
$C_7^{\mathit{eff}}$ and additionally on the coefficients of 
two four-Fermi operators, $C_{9}$ and $C_{10}$.
It has been argued \cite{AGM94} that the signs and
magnitudes of all three coefficients $C_7^{\mathit{eff}}$, $C_{9}$ and 
$C_{10}$
can, in principle,  be determined from the decays $\BGAMAXS$ and \bxsll 
~by measuring the dilepton invariant mass and Forward-Backward 
charged lepton asymmetry \cite{amm91}.

\par
 The SM-based rates for the decay \bsll , calculated in the free quark decay
approximation, have been known in the LO approximation \cite{BSGAM}
 for some time.
The NLO contribution reduces the scheme-dependence of the LO result in these
decays \cite{MisiakBM94}. In addition,
long-distance (LD) effects, which are expected to be very important in the
decay \bxsll, have also been estimated from data
 on the assumption that they arise dominantly due to
the charmonium resonances $ J/\psi,\psi',...$ through the decay chains
$B \rightarrow X_s J/\psi (\psi',...) \rightarrow X_s \ell^+ \ell^-$.
The resulting dilepton distribution is then a coherent sum of the resonating
(LD) and mildly varying (SD) contributions. Data near the resonances can be
used to better parametrize the LD contribution in future than is the case 
now. 
Recently, these LD-contributions have also been predicated upon assuming
that far from the resonant-region they can be determined in terms of the 
$1/m_c^2$ contributions in the HQET approach \cite{BIR97,Savage97}. This 
remains an interesting conjecture but impossible to test experimentally,
as, first of all, they represent just a class of power corrections and,
more importantly, this theoretical framework breaks down near the resonances.

 The leading $(1/{m_b}^2)$ power corrections
to the partonic decay rate and the dilepton invariant mass distribution
have been calculated  in the 
HQET approach \cite{FLS94}; these results  
have, however, not been confirmed in a recent 
independent calculation
\cite{AHHM96}, which finds that the power corrections in the branching
ratio ${\cal B}(B \to X_s \ell^+ \ell^-)$ are small (typically
$-1.5\%$). The corrections in the dilepton mass spectrum and the FB
asymmetry  are also small over a good part of this spectrum. 
However, 
the end-point dilepton invariant mass spectrum is not calculable in the
heavy quark expansion and will have to be modeled. As an 
alternative, non-perturbative 
effects in \bxsll have also been estimated \cite{AHHM96}, using the Fermi 
motion model discussed earlier
\cite{Aliqcd}. These effects, which model power corrections in $1/m_b$,  are 
also found to be small over most of the phase space except for the end-point
dilepton mass spectrum where they change the underlying parton model 
distributions significantly and have to be taken into account in
the analysis of data \cite{AHHM96}. 

Taking into account the spread in the values of the input parameters,
$\mu, ~\Lambda, ~\mt$, and ${\cal B}_{SL}$
discussed in the previous section in the context of ${\cal B}(B \to X_s + 
\gamma)$, the following branching ratios for the SD-piece have been estimated
\cite{AHHM96}:
\begin{eqnarray}\label{brbsll}
{\cal B}(\bxsee) &=& (8.4 \pm 2.3) \times 10^{-6}, \nonumber\\
{\cal B}(\bxsmm) &=& (5.7 \pm 1.2) \times 10^{-6}, \nonumber\\
{\cal B}(\bxstt) &=& (2.6 \pm 0.5) \times 10^{-7}, 
\end{eqnarray}
where theoretical errors and the error on ${\cal B}_{SL}$ have been added in 
quadrature.
 The experimental upper limit for the inclusive branching ratio for the decay
$\bxsmm$ was quoted by the UA1 collaboration some time
ago \cite{UA1R}, ${\cal B}(\bxsmm) < 5.0 \times 10^{-5}$. This limit has been
put to question in a recent CLEO paper \cite{CLEObsmm}.
From the CLEO data, a limit ${\cal B}(\bxsmm) < 5.8 \times 10^{-5}$ has
been set on this decay, with ${\cal B}(\bxsee) < 5.7 \times 10^{-5}$.
 Combining the di-electron and 
di-muon upper limits, CLEO quotes \cite{CLEObsmm} ${\cal B}(B \to X_s \ell^+ 
\ell^-) < 4.2 \times 10^{-5}$ (all limits are at $90\%$ C.L.). This is a 
factor 6 away from the SM 
estimates. As far as we know, there is no experimental limit on the mode
$X_s \tau^+ \tau^-$.
For a more detailed discussion of the modes $B \to X_s \ell^+ \ell^-$,
some related exclusive decays such as $B \to (K,K^*) \ell^+ \ell^-$, and
other rare $B$-decay modes, we refer to recent reviews 
\cite{ALI96,ALI97}. The CKM-suppressed decays $B \to X_d \ell^+ \ell^-$,
which are expected to be typically a factor 20 below their corresponding 
CKM-allowed decay rates, and their role in determining the CKM parameters 
have been discussed elsewhere \cite{KS196,Sanda97}.
\section*{Acknowledgments}
 I would like to thank  
Christoph Greub, Jim Smith, Tomasz Skwarnicki and Mark Williams for helpful
discussions. I also thank Mark for sending me a copy of the ALEPH paper
on $b \to s + \gamma$ submitted to the EPS conference. The warm 
hospitality of the organizing committee is greatly appreciated.

\end{document}